\documentclass[preprint,showpacs,preprintnumbers,amsmath,amssymb,aps]{revtex4-1}
\usepackage{graphicx,dcolumn,bm}% Include figure files
\usepackage{amssymb,amstext,amsmath}

\usepackage{graphicx}% Include figure files
\usepackage{dcolumn}% Align table columns on decimal point
\usepackage{bm}% bold math
\usepackage{amssymb}
\usepackage{multirow}
\usepackage{bigstrut}
\usepackage{makecell,rotating}

\usepackage{mathrsfs}
\usepackage{booktabs}
\usepackage{threeparttable}
\usepackage{multirow}
\usepackage{subfigure}
\usepackage{epsfig}
\usepackage{threeparttable}
\usepackage{chngpage}
\usepackage{float}

\usepackage{amstext}% bold math
\usepackage{amsmath}% bold math

\begin{document}

\title{Evolution of Cooperation on Temporal Networks}
\author{Aming Li$^{1,2,3}$, Lei Zhou$^{1,4}$, Qi Su$^{1}$, Sean P. Cornelius$^{2,5}$, \\Yang-Yu Liu$^{5,6,*}$, and Long Wang$^{1,}$}  
\email{Corresponding authors.}
\affiliation{
$^{1}$Center for Systems and Control, College of Engineering, Peking University, Beijing 100871, China \\
$^{2}$Center for Complex Network Research and Department of Physics, Northeastern University, Boston, Massachusetts 02115, USA \\
$^{3}$Department of Physics, Physics of Living Systems Group, Massachusetts Institute of Technology, Cambridge, Massachusetts 02139, USA \\
$^{4}$Department of Ecology and Evolutionary Biology, Princeton University, Princeton, New Jersey 08544, USA \\
%08544-1003
$^{5}$Channing Division of Network Medicine, Brigham and Women's Hospital, Harvard Medical School, Boston, Massachusetts 02115, USA \\
$^{6}$Center for Cancer Systems Biology, Dana-Farber Cancer Institute, Boston, Massachusetts 02115, USA
}
\date{\today}

% 突出时变网络的特点和其广泛性
\begin{abstract}

The structure of social networks is a key determinant in fostering cooperation and other altruistic behavior among naturally selfish individuals. 
However, most real social interactions are temporal, being both finite in duration and spread out over time.
This raises the question of whether stable cooperation can form despite an intrinsically fragmented social fabric.
%This raises the question: How does the temporality of social interactions affect the evolution of cooperation?
Here we develop a framework to study the evolution of cooperation on temporal networks in the setting of the classic Prisoner's Dilemma.
By analyzing both real and synthetic datasets, we find that temporal networks generally facilitate the evolution of cooperation compared to their static counterparts.
More interestingly, we find that the intrinsic human interactive pattern like bursty behavior impedes the evolution of cooperation.
Finally, we introduce a measure to quantify the temporality present in networks and demonstrate 
that there is an intermediate level of temporality that boosts cooperation most.
%In other words, both high and low temporality relatively help the spreading of egoistic behavior on temporal networks.
Our results open a new avenue for investigating the evolution of cooperation in 
more realistic structured populations.
\end{abstract}
\maketitle

\section{Introduction}
Understanding and sustaining the evolution of cooperation in human and animal societies have long been a challenge since Darwin \cite{Hamilton63AN,Trivers71QRB,Smith76QRB,Hofbauer98book,Nowak06book}.
Evolutionary game theory offers a prominent paradigm to explain the emergence and persistence of cooperation among egoists,
and many results have been obtained from analytical calculations \cite{Hofbauer98book,Nowak06book}, numerical simulations \cite{Nowak06book,Szabo2007}, and behavioral experiments \cite{Traulsen2010,Grujic10PLoSONE,Suri11PLoSONE,Rand2011,Yamir2012,Wang2012,Rand2014,Antonioni2015}.
Traditionally, researchers have been focusing on the well-mixed or homogeneous population scenarios \cite{Trivers71QRB,Smith76QRB,Hofbauer98book,Nowak92Nature}. 
Yet, both spatial 
population structures and social networks suggest that real populations are typically not well-mixed.
Indeed, in a population some individuals may interact more likely than others do.
In both theory and experiments, the ideally well-mixed scenario has been extended to heterogeneous structured populations represented by complex networks, where nodes represent individuals and links capture who interacts with whom \cite{Nowak92Nature,Lieberman2005a,Santos05PRL,Yamir2012,Rand2014}.
And a unifying framework coined as network reciprocity is proposed for emergence of cooperation in structured populations \cite{Nowak2006}, especially for the networks with the degree heterogeneity which is typically observed in scale-free networks \cite{Santos05PRL,Santos06PNAS}.

Despite their deep insights, those works all rely on a key assumption that the contact graph or the interaction network of individuals is time invariant. 
In reality, this assumption is often violated, especially in social networks, 
where contacts between individuals are typically short-lived.
Emails and text messages for example represent near-instantaneous and hence ephemeral links in a network. 
Even in cases where the contacts have non-negligible durations --- such as phone calls, or the face-to-face interactions between inpatients in the same hospital ward --- their finite nature means that the network structure is in constant flux.
It has been shown that the temporality of edge activations can noticeably affect various dynamical processes, ranging from
the information or epidemics spreading \cite{Slowingd13PRL,Scholtes2014natcommun,BurstTres13PloSone,Slowingdon12PRE} to network accessibility \cite{Accessibility13PRL} to controllability \cite{Li2016b}.

It is natural to expect that temporality will have a similarly profound effect in social systems, particularly in situations when individuals engage in interactive behavior.
Indeed, if Alice interacts with Bob who only later betrays Charlie, Alice's behavior toward Bob could not have been influenced by his later treachery.
Yet the links A---B---C would be ever-present in a static representation of this social network.
Despite some existing efforts \cite{Cardillo2014}, up to our knowledge, the impact of temporal networks on the evolution of cooperation has not been systematically explored. 
It is still unclear whether the temporality will enhance the cooperation or not.

Here for the first time, we explore the impacts of temporality of human interactions on the evolution of cooperation over both empirical and synthetic networks.
Moreover, the impacts of the bursty behavior rooted in human activity on the evolution of cooperation are also investigated.

\section{Model}
We conduct our investigation in the setting of classic evolutionary game theory, in which each of two players may choose a strategy of cooperation (C) or defection (D).
Each receives a payoff $R$ for mutual cooperation, and an amount $P$ for mutual defection.
When the players' strategies disagree, the defector receives a payoff $T$ while the cooperator receives $S$.
These outcomes can be neatly encoded in the payoff matrix
\begin{equation*}
\bordermatrix{%
& C & D  \cr
C & R & S  \cr
D & T & P \cr
}
\end{equation*}
where the entries give the payoff each player receives under different combinations of strategy.
For simplicity, we focus on the case where $R=1, T=b$ and $S=P=0$, leaving the sole parameter $b>1$, which represents the temptation to defect and hence the system's tendency toward selfish behavior.
This parameter choice corresponds to the classic
Prisoner's Dilemma, wherein the optimal strategy for any single individual is to defect, while mutual cooperation is the best choice for the alliance \cite{Nowak92Nature,Santos05PRL,Vainstein07JTB,Gomez2007}.

\begin{figure}[H]
\centering
\includegraphics[width=\textwidth]{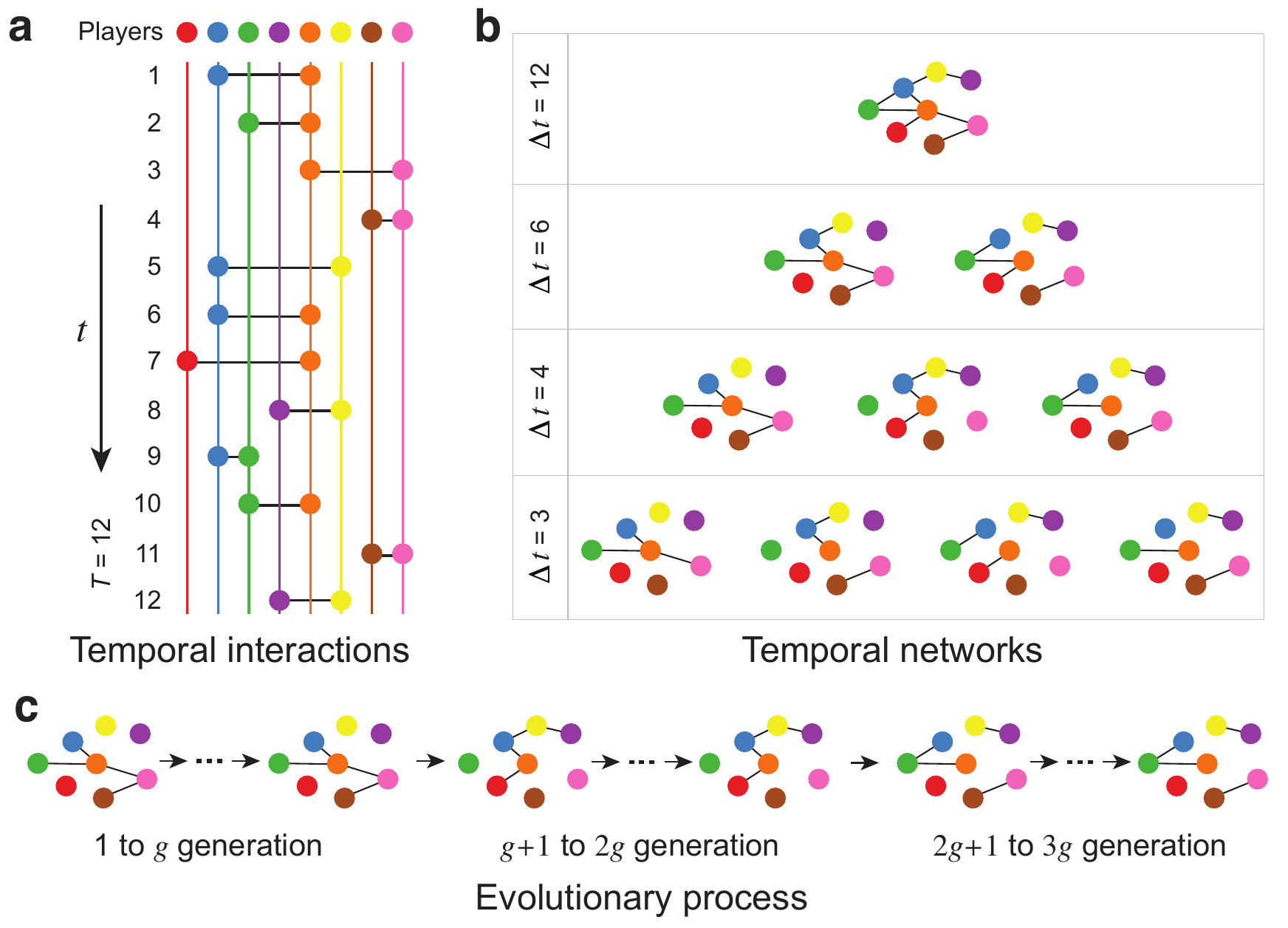}%
\caption{
\textbf{Construction of temporal networks from temporal interactions, and evolutionary process on temporal networks.}
\textbf{(a)} Temporal interactions between $8$ individuals indicated by solid circles with different colors.
Along the whole time from $t=1$ to $t=T$, each individual is depicted by the same color line, over which the corresponding circles will be given and connected with each other at time $t$ provided two players interact with each other during the time interval $(t-\tau,t]$. 
Here $\tau=1$ for the simplicity of visualizations, and normally in the real data collected by SocioPatterns \cite{datawebsite}, $\tau=20$s.
\textbf{(b)} 
Four different temporal networks that arise from aggregating the interactions shown in (a) into snapshots using different time windows $\Delta t$.
When $\Delta t = T$, all interactions are captured in a single snapshot, corresponding to the static network that is the typical object of study in social network data.
In general, when $\Delta t<T$, we have $\lceil T/ \Delta t \rceil$ snapshots.
\textbf{(c)} The definition of evolutionary process on temporal networks.
Taking the temporal network corresponding to $\Delta t=4$ in (b) as an example, we perform $g$ generations of evolution in each snapshot before changing the network structure to the next one, and totally we run $G$ generations until the composition of the population is stable.
If $\lceil T/ \Delta t \rceil * g < G$, 
we repeat the sequence of snapshots from the beginning until convergence.
}
\label{fig_1}
\end{figure}

Figure~\ref{fig_1} illustrates the essence of our framework.
We consider a temporal network 
to be a sequence of separate networks on the same set of $N$ nodes, which we call \emph{snapshots}. 
These snapshots are constructed from empirical data by aggregating social contacts over successive windows of $\Delta t$ (Fig.~\ref{fig_1}a and~\ref{fig_1}b), yielding the links active in that snapshot.
To capture the interactions occurring on these networks,
we initially set an equal fraction of cooperators and defectors (network nodes) in the population
on the first snapshot.
At the beginning of each generation  (round of games), every individual $i$ plays the above game with each of its $k_i$ neighbors, accumulating a total payoff $P_i$ according to the matrix above.
At the end of each generation each player $i$ may change his or her strategy, by randomly picking a neighbor $j$ with payoff $P_j$ from
its $k_i$ neighbors, and then imitating $j$'s strategy with probability $(P_j-P_i)/(Dk_d)$ if $P_j > P_i$.
Here $D=T-S$ and $k_d$ is the larger of $k_i$ and $k_j$ \cite{Santos05PRL,Meloni2009ContSpac}.
We repeat this procedure $g$ times within each snapshot before changing the network structure (Fig.~\ref{fig_1}c). 
In this way, $g$ controls the timescale difference between the dynamics on the network and the dynamics of the network.
We continue running the game, changing the network structure every $g$ generations, until the system reaches a stable fraction of cooperators, $f_c$.

\section{Results}
\subsection{Temporal networks facilitate the evolution of cooperation}
\label{part1}

\begin{figure}[H]
\centering
\includegraphics[width=\textwidth]{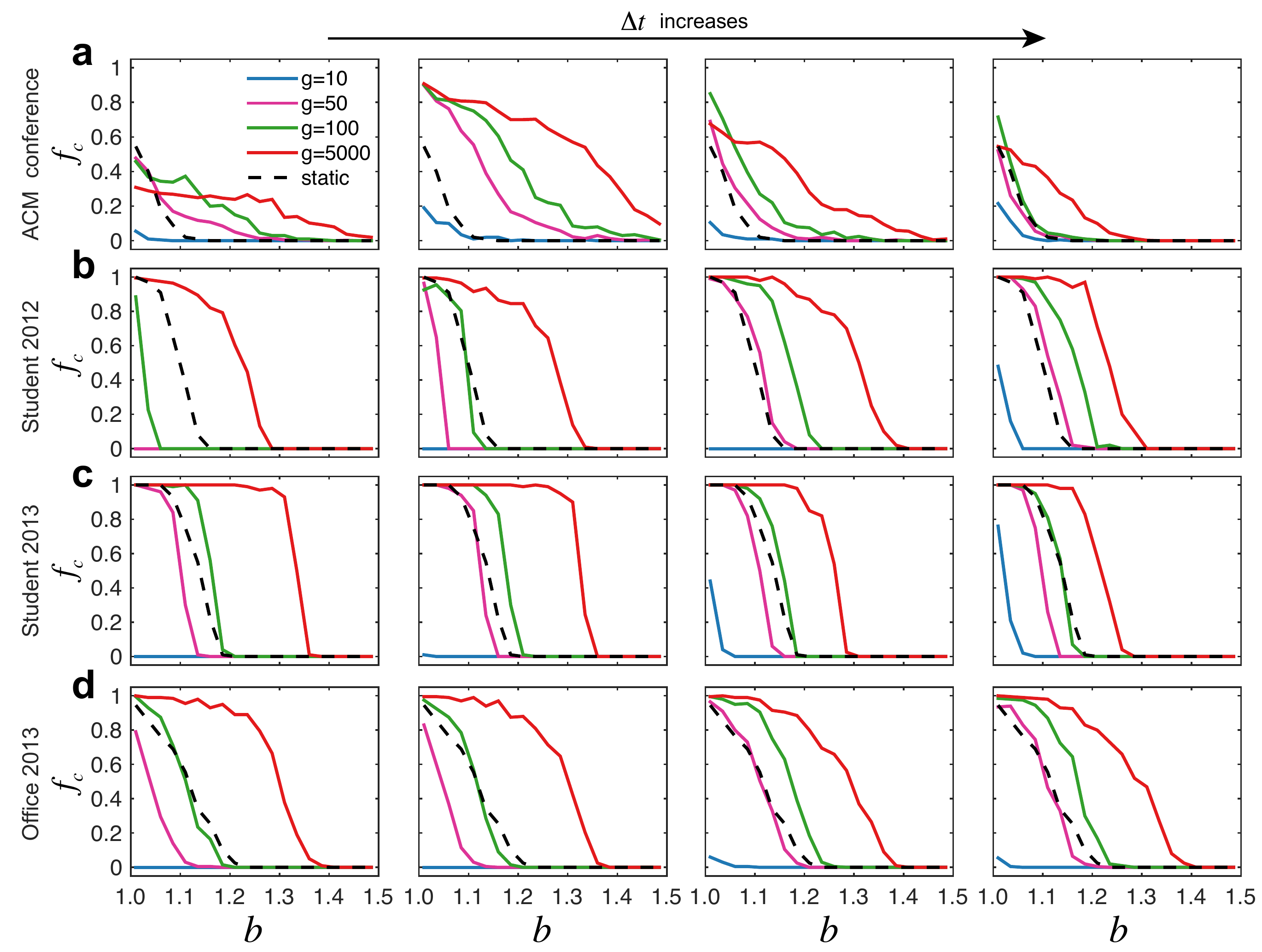}%
\caption{
\textbf{Temporal networks generally promote the evolution of cooperation in real social systems.}
For four empirical datasets: \textbf{(a)} the ACM conference, \textbf{(b)} Student 2012, \textbf{(c)} Student 2013, and \textbf{(d)} Office 2013, we show the stable frequency of cooperation on both temporal (colored lines) and static (black dashed lines) networks with different values of the aggregation time windows $\Delta t$.
We choose $1,2,6,24$ hours from left to right in (a) to (c) and $6,8,12,24$ hours in (d), respectively.
After letting the population evolve $g$ generations on each snapshot, we average over another $2,000$ generations after $G=10^6$ generations on each temporal network, to obtain the stable frequency of cooperators.
The statistics of each dataset are given in Table~\ref{fig_tabel}.
}
\label{fig_2}
\end{figure}

Our principal result is the temporal networks generally enhance cooperation relative to their static counterparts, and allow it to persist at higher levels of temptation $b$. 
Figure~\ref{fig_2} shows the equilibrium fraction of cooperators $f_c$ for temporal networks formed from empirical data of four social systems \cite{datawebsite}: attendees at a scientific conference (ACM conference) \cite{Isella2011}, students at a high school in Marseilles, France in two different years (Student 2012 \cite{Fournet2014}, Student 2013 \cite{Mastrandrea2015School2013}), and workers in an office building (Office 2013) \cite{Genois2015OfficeData}. 
In each of these systems there exists a broad range of $g$ over which $f_c$ is greater in the temporal network than in its static equivalent, at almost all values of $b$. 
This is true even for small $\Delta t$. 
Here the network's links are distributed over a large number of rarefied snapshots, leaving little network ``scaffolding'' on which to build a stable cooperation. 
Nonetheless, there exists a range of $g$ that can compensate for this sparsity, again leaving temporal networks the victor. Indeed, the only situation in which temporal networks are less amenable to cooperation than static networks is when $g$ is small. 
In this limit, the evolutionary timescale is comparable to the dynamical timescale, and patterns of cooperation have no time to stabilize before being disrupted by the next change in network structure.

\begin{figure}[H]
\centering
\includegraphics[width=\textwidth]{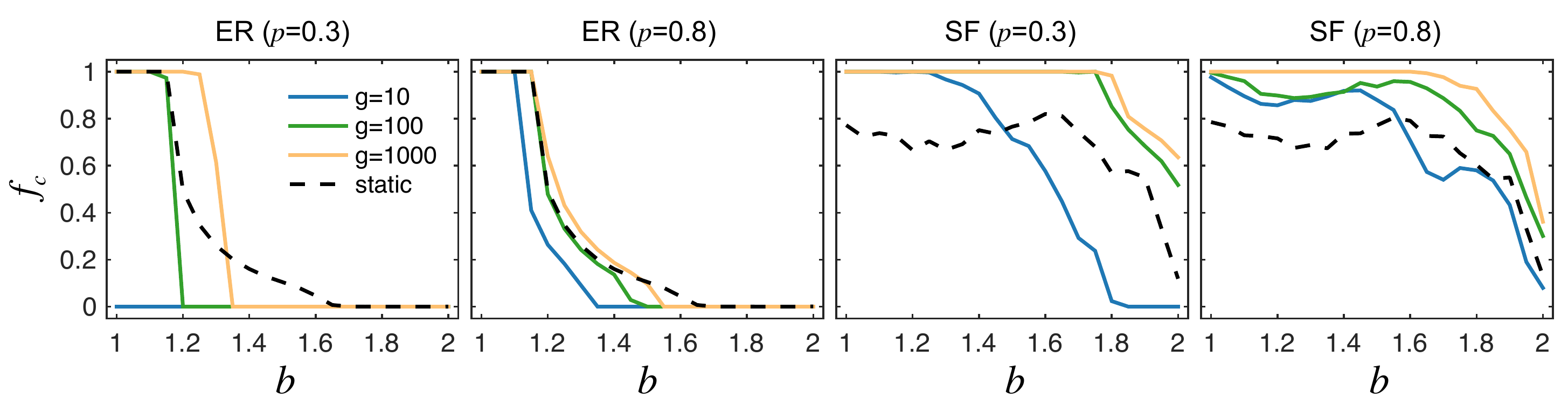}%
\caption{
\textbf{Evolution of cooperation on temporal networks generated from synthetic data.
}
Here we generate $M$ sparse snapshots based on a base 
scale-free network formed from static model \cite{Goh2001}, and a base Erd\H{o}s-R\'enyi random \cite{ER1960} network, choosing a fraction $p$ of edges to be active within each snapshot.
Here $M=100$, the network size $N=1000$, and average degree $\langle k \rangle=10$.
The robustness of the corresponding results for other parameters and other methods of generating synthetic temporal networks has been verified (see Fig.~\ref{fig_S_synthetic} in the SI).
}
\label{fig_3}
\end{figure}

To test whether these results arise from the specific temporal patterns in real social systems, we have also simulated games on temporal versions of synthetic scale-free (SF) \cite{Barabasi1999a} and Erd\H{o}s-R\'enyi (ER) \cite{ER1960} random networks (see Methods). 
We again find that with almost level of temporality (\emph{i.e.}, $g < \infty$), cooperators have an easier time gaining footholds in the population (Fig.~\ref{fig_3}). 
Interestingly, the temporal scale free networks yield a higher $f_c$, all other things being equal, than the temporal ER networks (Fig.~\ref{fig_3} and Fig.~\ref{fig_S_synthetic}). 
As such, temporality preserves the cooperative advantage of heterogenous populations, previous observed in static networks \cite{Santos05PRL}.

\subsection{Effects of burstiness on the evolution of cooperation}
Analyses of the temporal patterns of human interactions in email \cite{Karsai11PRE}, phone calls \cite{Karsai11PRE}, and written correspondence \cite{Barabasi05} have revealed a high degree of burstiness --- periods of intense
activity followed by ``lulls'' of relative silence. 
Such correlations embedded in temporal interactions have been shown to have effects on network dynamics above and beyond those of temporality alone, for instance accelerating the spread of contagions \cite{Rocha2011,BurstTres13PloSone}.
We have verified that burstiness is present to varying degrees in the four data sets we study, in the form of a power law distribution of inter-event times between the node activations (Fig.~\ref{fig_S_BurstGama}). 
But to what extent do these patterns help or hinder the evolution of cooperation?

\begin{figure}[H]
\centering
\includegraphics[width=\textwidth]{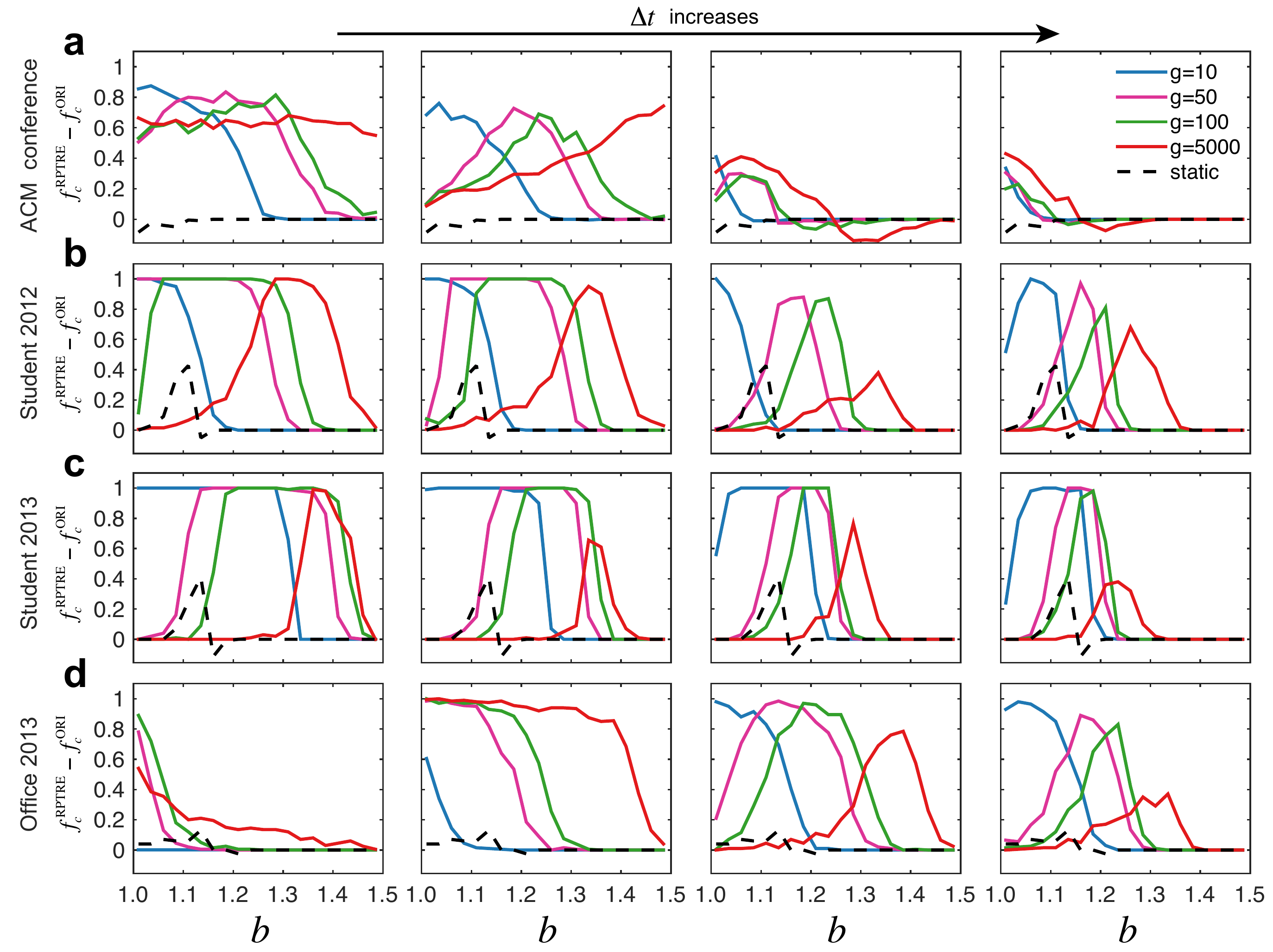}%
\caption{
\textbf{The intrinsic bursty behavior in human interactions suppresses the maintenance of cooperation.}
For each dataset, we show the difference $f_c^{\text{RPTRE}} - f_c^{\text{ORI}}$ between the stable frequency of cooperators $f_c^{\text{RPTRE}}$ in temporal networks generated from each dataset after 
randomly permuting both the timestamps and edges (RPTRE in the Methods)
which 
erases the burstiness inherent to human interaction data
(see Methods), and $f_c^{\text{ORI}}$ over the original scenarios.
We see that the frequency of cooperators generally increases after the bursty behavior is destroyed, 
suggesting that 
correlations in activity within a social network
is antagonistic toward the formation of cooperation.
Results on each dataset after randomizations with different null models \cite{Holme2012} are given in Figs.~\ref{fig_S_ht09} to~\ref{fig_S_Office2013} in the SI.
Other parameters are the same as those in Fig.~\ref{fig_2}.
}
\label{fig_4}
\end{figure}

We have studied this question by randomizing the contacts in each of the datasets we study, both their source and target $(i, j)$ and their timestamps $t$. %%%%%%
We stress that this randomization has the effect of erasing bursty behavior at the level of individual node. 
In every temporal network, we find that cooperation is improved when the natural burstiness is removed in this way, 
suggesting that bursty behavior impedes the evolution of cooperation (Fig.~\ref{fig_4}). %%%%% correlations, link activity...
For the effects of other null models that permute only the structure or the time stamps of the contacts, please refer to Figs.~\ref{fig_S_ht09} to~\ref{fig_S_Office2013}, where we also show that the above results are robust after the data is randomized with various methods.  %%%%%
Furthermore, this is true for nearly all choices of parameters $\Delta t$, $g$, and $b$. 
But how do we make sense of these findings in relation to the observation above, namely that real temporal networks generically promote cooperation?  %%% combination for the next paragraph

\subsection{Temporality determines the fate of cooperators}
The parameters $g$ and $\Delta t$, and the burstiness represent three different facets of temporality. 
Specifically, the relationship between the dynamical/structural timescales, the amount the network structure is spread over time, and the correlations between the associated snapshots, respectively. 
To understand the effects of these parameters in a unified way, we introduce the following measure of the \emph{temporality} $\mathcal{T}$ of a temporal network with $M$ snapshots as %%% 
$$\mathcal{T} = \frac{1}{M-1} \sum_{m=1}^{M-1} \frac{\sum_{i,j}|a_{ij}(m) - a_{ij}(m+1)|}{\sum_{i,j}\max\{a_{ij}(m),a_{ij}(m+1)\}}.$$
Here $a_{ij}(m)$ is the connectivity between nodes $i$ and $j$ in snapshot $m$, being $1$ if the nodes have a contact in the associated time window and $0$ otherwise, and the above fraction equals to $0$ for any two nearby empty networks without links. 
This measure captures the likelihood that any currently inactive link will become active in the next snapshot (or conversely, that an active link becomes inactive).
If we need to replay the temporal network $M$ is $\lceil T/ \Delta t \rceil$, and $\lceil T/ \Delta t \rceil -1$ if we do not. %%%
For a temporal network, generally $0<\mathcal{T} \leq 1$, and $\mathcal{T} = 0$ for static network where network topology does not change with time. %%%%
	
\begin{figure}[H]
\centering
\includegraphics[width=\textwidth]{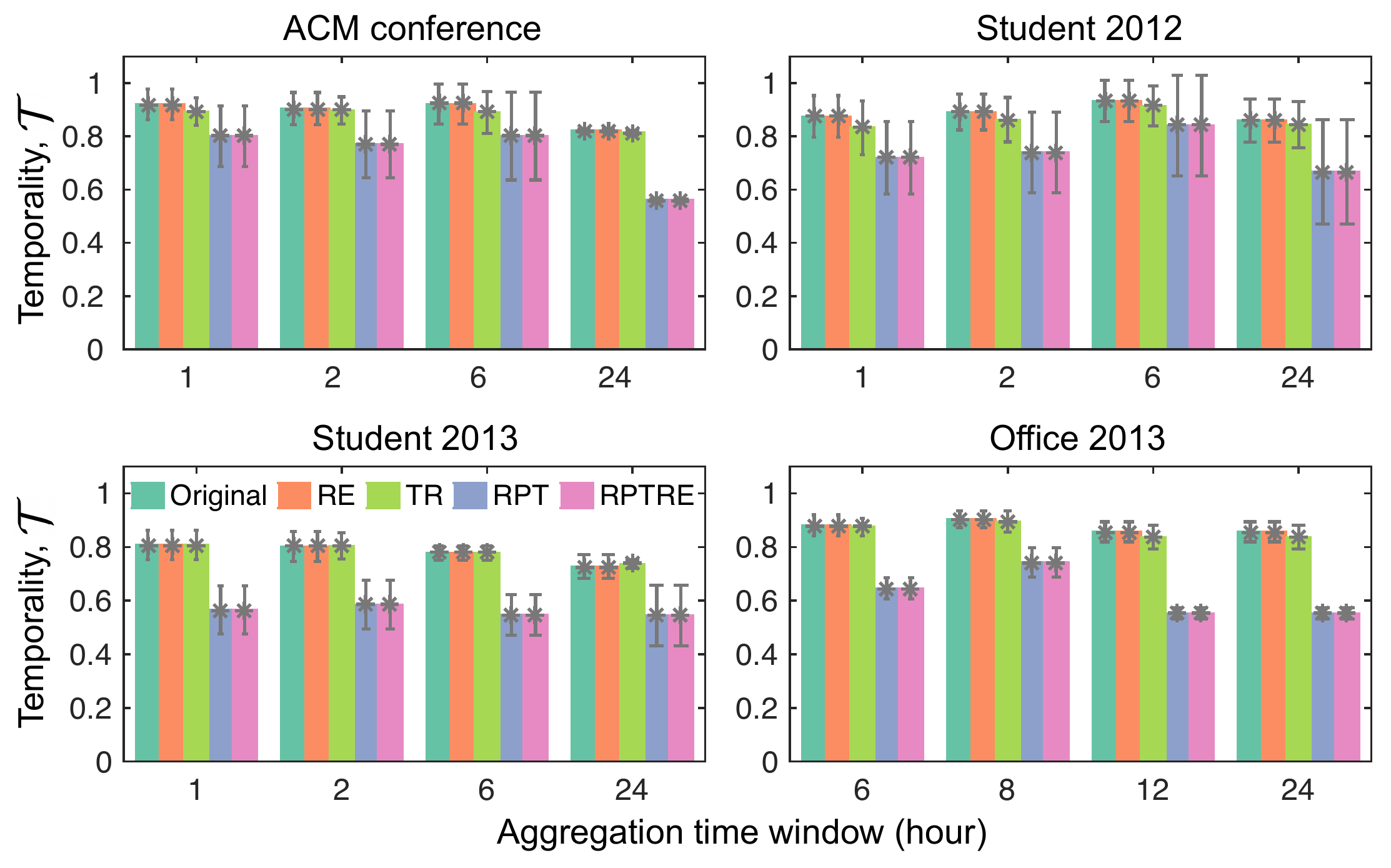}%
\caption{
\textbf{The temporality of real networks.}
The temporality $\mathcal{T}$ of the original datasets is shown alongside their randomizations for different time windows $\Delta t$.
We find that by destroying burstiness, randomizations altering the time ordering of contacts (RPT, RPTRE) decrease $\mathcal{T}$.
Figure~\ref{fig_S_DetailT} shows how the overall temporality $\mathcal{T}$ changes on a snapshot-by-snapshot basis, and the corresponding standard deviation is given in this figure.
}
\label{fig_5}
\end{figure}

Figure~\ref{fig_5} shows the value of $\mathcal{T}$ computed for the original and randomized versions of each of the four data sets we study. 
We see that the original data displays high temporality, which decreases following the randomization procedure (RPTRE) described above. 
Considering that $f_c$ for the randomized temporal networks is typically higher than in the originals (Fig.~\ref{fig_4}), this suggests that too high temporality is inimical to the spread of cooperation, instead fostering egoistic behavior. 
On the other hand, too low of a $\mathcal{T}$ is also associated with diminished cooperation. 
For example, $f_c$ is not maximal in Fig.~\ref{fig_2} for $\Delta t=24$, which corresponds to snapshots that are relatively dense and slowly changing, paving the way for defectors to extort cooperators. 
Altogether, the picture that emerges is one of an intermediate regime --- a ``sweet spot'' of temporality in which cooperation is enhanced relative to static systems.

\section{Conclusion and Discussion}

Considering the real characteristics of human interactions where the underlying networks are temporal and possess the underlying interactive patterns, we have addressed the evolution of cooperation on temporal networks.
After finding that temporal networks from empirical datasets favor the evolution of cooperation more than their static counterparts, we also validate our results on synthetic networks.
This central finding holds even after the empirical data is randomized, 
thereby destroying specific temporal patterns (such as bursts) characterizing real human interactions.
Altogether, this suggests that temporality --- and temporality alone --- is sufficient to improve cooperation.
Interestingly, after randomizations, we find that the level of cooperation is further improved suggesting that the bursty nature of human interactions hinders the maintenance of cooperation to some degree.
At last, we demonstrate that the temporality of a temporal network determines the fate of cooperators, with cooperators flourishing at intermediate values of the network temporality.
By virtue of both empirical and traditional synthetic data, our explorations systematically illustrate the effects of temporality  on the evolution of cooperation.

Note that the intrinsic temporal nature of the contact graph or interaction network is fundamentally different from the slight change of population structure due to individuals' migration \cite{Helbing09PNAS,Roca11PNAS,Li2013a}.
% as individuals pursue high payoffs \cite{Helbing09PNAS} or are not satisfied with their current circumstances \cite{Roca11PNAS,Li2013a}.
The latter is usually restricted to the elaborate rules or strategies 
based on a presumed synthetic static network \cite{Aktipis2004,Vainstein07JTB,Helbing09PNAS,Meloni2009ContSpac,BWu10PLoSONE,Roca11PNAS,Wu2012,Li2013a,Antonioni2014,Tomassini2015,Pinheiro2016}. 
%such as stochastic \cite{Vainstein07JTB,Meloni2009ContSpac,BWu10PLoSONE,Antonioni2014}, success-driven \cite{Helbing09PNAS}, or expectation-driven migration \cite{Roca11PNAS,Wu2012,Li2013a}. 
The coevolution of the network and strategy has been studied in the case where the network changes passively and with small temporality under constant average degree and population size \cite{Skyrms2000,Zimmermann2004,Santos06PLoSCB,Szolnoki08EPL,FF08PRE,Perc2010}.
These coevolutionary dynamics arise from players' strategic switch of partners, 
a process typically governed by pre-determined mechanisms.
%which is determined by pre-designed mechanisms essentially.
% and the intrinsic temporal human interactions cannot be reflected by the passive evolution of population structures solely.
However, it is unlikely that the natural temporality observed in real human social dynamics is driven exclusively (or even primarily) by strategic switching in pursuit of a given objective.
% although it is also called as dynamic network sometimes.
%Although the real dataset is noticed in \cite{Cardillo2014}, only a single time window is used to aggregate dynamic network, which restricts the number of snapshots as a constant, and also individuals are imposed to update their strategies infrequently, leading to a result that dynamic network favors selfish behavior. 

Another natural extension of the current work on temporal networks is to consider the group interactions, which involve the interactions between individuals who are not directly connected with one another \cite{Santos08Nature,Szolnoki09PRE,Li2016,Zhou2015}.
These interactions generate much more dynamical complexity, which cannot be captured by pairwise interactions \cite{Gokhale2010,Li2015JTB}.
This is also true in microbial populations, where even pairwise outcomes could 
%been shown the power to predict 
predict
the survival of three-species competitions with high accuracy, yet information from the outcomes of three-species competitions is still needed as we want to predict the scenario over more number of species \cite{Jeff2016}.  %%%%
Moreover, the menu of strategies can be expanded beyond the simple dichotomy of cooperation versus defection.
For example, the canonical three strategies game rock-paper-scissors, which may serve as a model to study the biological diversity in microbial populations and communities \cite{RPS1996,Durrett1997,RPS2002}.

Finally, our results have implications for other dynamical processes occurring on temporal networks.
If we regard the evolution of cooperation on temporal networks as a spreading dynamics of different strategies, it
may serve as a new angle to investigate other related dynamics.
For example, consider epidemic spreading, where the temporal network characteristics of networks had been shown to either speed up \cite{EpidemicS2011SpeedUp,BurstTres13PloSone} or slow down \cite{Karsai11PRE,ESpr2011SlowDown} the spreading, and the shuffle of time stamps was shown to enhance the spreading in a network of sex buyers and prostitutes \cite{Rocha2011}. %%%
After evaluating the payoffs (benefits and costs) of susceptible and infected individuals as they encounter one another, our framework of the evolution of cooperation may help us understand more phenomena including the epidemic spreading on temporal networks.

\section*{Methods}
\textbf{Empirical temporal networks.}
We construct temporal networks from empirical datasets \cite{datawebsite} by aggregating contacts into undirected network links over time windows of $\Delta t$ (Fig.~\ref{fig_1}a). 
Here, a contact is a triplet $(t, i, j)$ representing the fact that individuals $i$ and $j$ interacted during the time interval $(t, t+20\text{s}]$.
In this way, we obtain a temporal network with $\lceil T/ \Delta t\rceil$ snapshots, where $T$ is the total time span of the dataset and $\lceil z \rceil$ 
is the smallest integer greater than or equal to $z$.
Thus the active time interval for the snapshot $m$ is from $(m-1)\Delta t$ to $m \Delta t$, and a link between $i$ and $j$ exists if players $i$ and $j$ interact at least once in that time period (Fig.~\ref{fig_1}b).
We obtain a static network in the limit where $\Delta t = T$.

\textbf{Synthetic temporal networks.}
We generate temporal versions of scale-free and random networks with size $N$ and average degree $\langle k \rangle$
by first generating a base static network, using static model \cite{Goh2001} and the Erd\H{o}s-R\'enyi model \cite{ER1960}, respectively.
We then form $M$ snapshots by randomly and independently choosing a fraction 
$p$ of edges to be active in each one.
We have verified that our results hold under more sophisticated methods for building temporal networks from a static network backbone, such as the activity-drive model \cite{Perra2012}

%\noindent
\textbf{Randomizations of empirical datasets.}
We consider four widely-used null models \cite{Holme2012} to randomize the empirical data:
%The null models \cite{Holme2012} we adopted to randomize the empirical data are:
Randomized Edges (RE) where we randomly choose pairs of edges $(i,j)$ and $(i',j')$, and replace them with $(i,i')$ and $(j,j')$ or $(i,j')$ and $(j,i')$ with equal probability provided this results in no self loops;
Randomly Permuted Times (RPT), where we shuffle the timestamps of the contacts, 
leaving their sources and targets unaltered;
%and leave the sources and targets of the links unaltered;
Randomly Permuted Times + Randomized Edges (RPTRE) which consists 
first of RPT followed by RE;
%of RPT followed by RE, where we first do RPT, and then RE;
and Time Reversal (TR), where the temporal order of the contacts is reversed.

\section*{Acknowledgments}
A.L., L.Z., Q.S. and L.W. are supported by NSFC (Grants No.~61375120 and No.~61533001).
A.L. and L.Z. also acknowledge the support from China Scholarship Council (CSC) with No.~201406010195 and No.~201606010270, respectively.
S.P.C. is supported by the John Templeton Foundation (No.~51977) and NIH (No.~P50HG004233).
Y.-Y.L. is supported by the John Templeton Foundation: Mathematical and Physical Sciences grant number PFI-777.

%\section*{Author Contributions}
%A.L., S.P.C., Y.-Y.L. and L.W. conceived and designed the project.  
%All authors performed the research. 
%A.L. and L.Z. performed calculations on the empirical data. 
%A.L. and Q.S. performed calculations on the synthetic data. 
%A.L., L.Z., Y.-Y.L. and L.W. analyzed the results.
%A.L., S.P.C., Y.-Y.L. and L.W. led the writing of the manuscript,
%L.Z. and Q.S. edited the manuscript. 
%

\newpage

\begin{table}[H] 
\caption{
\textbf{Statistics of the datasets.}
The four datasets we employed are interactions between: attendees of a ACM Hypertext conference over about 2.5 days from 8am on Jun.~29th 2009 (ACM conference), 
students in 5 classes 
at a high school in Marseilles, France over a period of 7 days in Nov.~2012 (Student 2012),
in 9 classes at a high school in Marseilles, France over 5 days in Dec.~2013  (Student 2013), 
individuals in an office building in France, from Jun.~24 to Jul.~3, 2013
(Office 2013). 
The number of snapshots is calculated based on the total time window $T$ over which the data were
collected, and $\Delta t$ 
(in seconds) is the time window used to aggregate the contacts into snapshots.
Contacts are defined as individual triples $(t, i, j)$ in the data, meaning nodes $i$ and $j$ were observed interacting in the time interval $(t,t+20\text{s}]$. 
Events (links), on the other hand, are continuous interactions formed by coalescing time-adjacent contacts between the same $i$ and $j$.
}
\center
\begin{tabular}{ccccc} 
\hline \hline
  \multirow{2}*{} &   ACM  conference &   Student 2012 &  Student 2013  & Office 2013\\ \hline
Number of nodes & $113$ & $180$ & $327$ & $92$  \\
Number of snapshots & $\lceil 212,360\text{s}/\Delta t \rceil$ & $\lceil 729,520\text{s}/\Delta t \rceil$ & $\lceil 363,580\text{s}/\Delta t \rceil$  & $\lceil 987,640\text{s}/\Delta t \rceil$ \\
Number of contacts & $20,808$ & $45,047$ & $188,508$ & $9,827$  \\
Number of events (links) & $9,865$ & $19,774$ & $67,613$ & $4,592$  \\
Recording period (day) & $2.5$ & $7$ & $5$  & $14$\\
Time resolution (second) & $20$    &   $20$    &   $20$  &   $20$ \\
\hline \hline 
\end{tabular}
\label{fig_tabel}
\end{table}

%\bibliography{../Public/bib_gametemp}

\newpage
\setcounter{figure}{0}
\renewcommand{\thefigure}{S\arabic{figure}}
\textbf{SUPPLEMENTARY INFORMATION (SI)}

\begin{figure}[H]
\centering
\includegraphics[width=\textwidth]{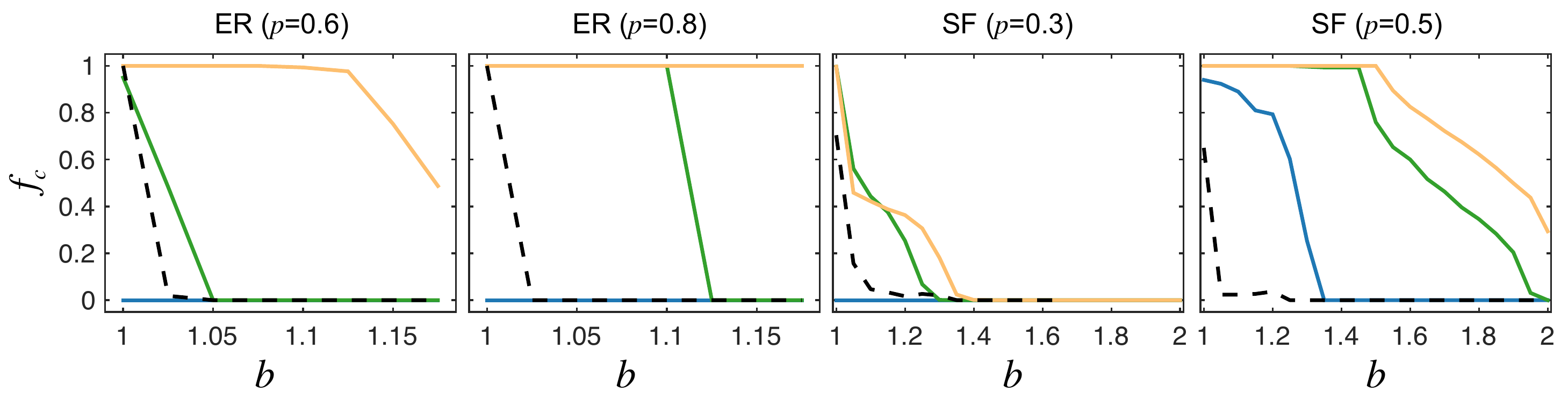}%
\caption{
\textbf{Evolution of cooperation on synthetic temporal networks.}
Here we generate $M$ sparse snapshots based on $M$ different 
scale-free networks with preferential attachment \cite{Barabasi1999a} and Erd\H{o}s-R\'enyi random networks \cite{ER1960} with the network size $N=1000$ and average degree $\langle k \rangle=4$ (see Methods).
Our results shown in Fig.~\ref{fig_3} are also validated with different $p$, which determines the level of link activity of each snapshot.
Other parameters are the same as those in Fig.~\ref{fig_3}.
}
\label{fig_S_synthetic}
\end{figure}

\begin{figure}[H]
\centering
\includegraphics[width=\textwidth]{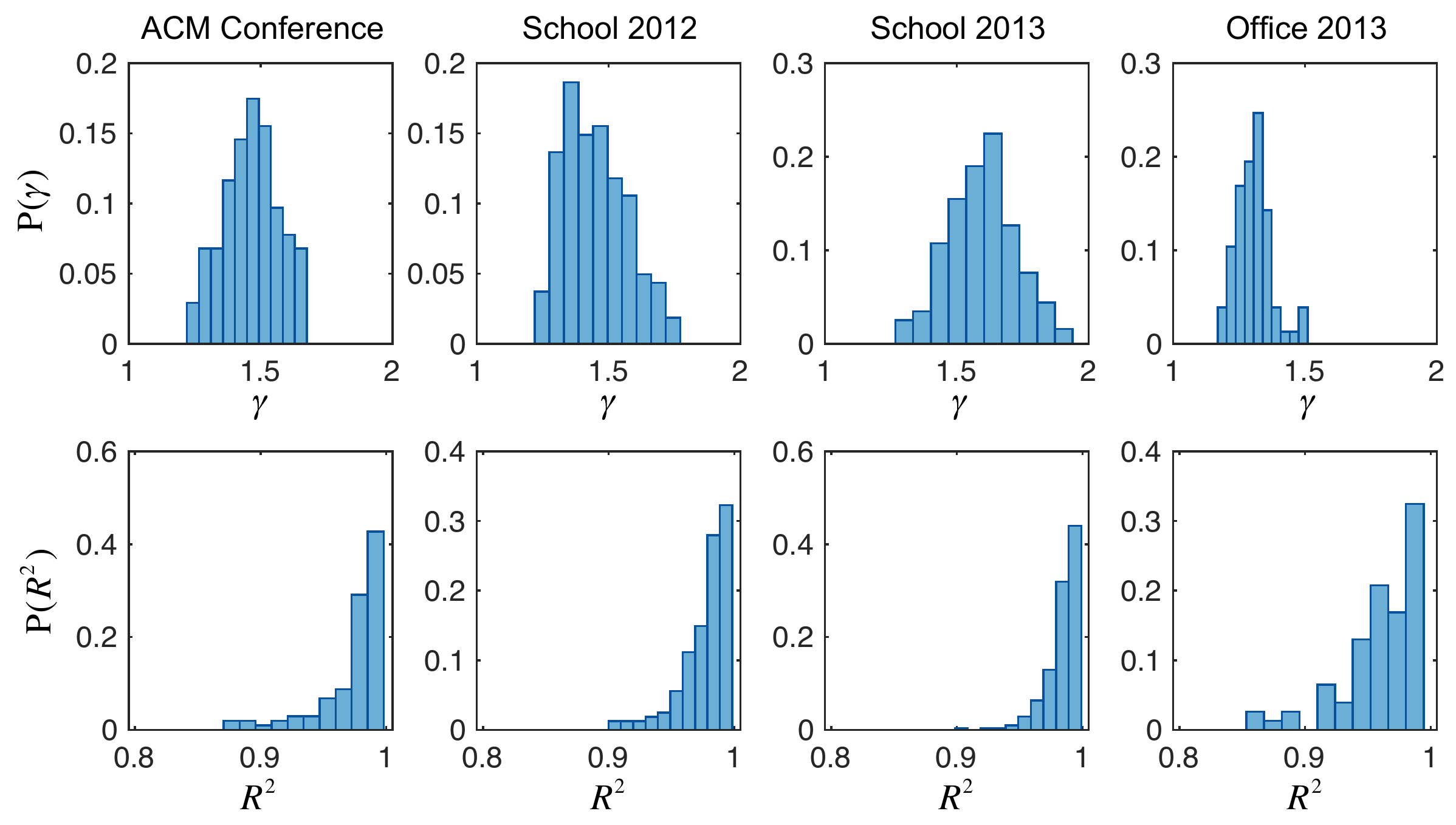}%
\caption{
\textbf{Bursty behavior in four datasets.}
For every dataset, we obtain a number of inter-event time $\tau$ for each individual based on his or her interactive logs.
For an individual, as the number is bigger than $30$, we fit all $\tau$ with power-law distribution $P(\tau) \sim \tau^{-\gamma}$, generating a $\gamma$ and an adjusted $R^2$ \cite{adjustedR}.
For each dataset, we give the distributions of $\gamma$ and the adjusted $R^2$ for all individuals there.
The second row shows that there are intrinsic bursty behavior in every original dataset.
}
\label{fig_S_BurstGama}
\end{figure}

\begin{figure}[H]
\centering
\includegraphics[width=\textwidth]{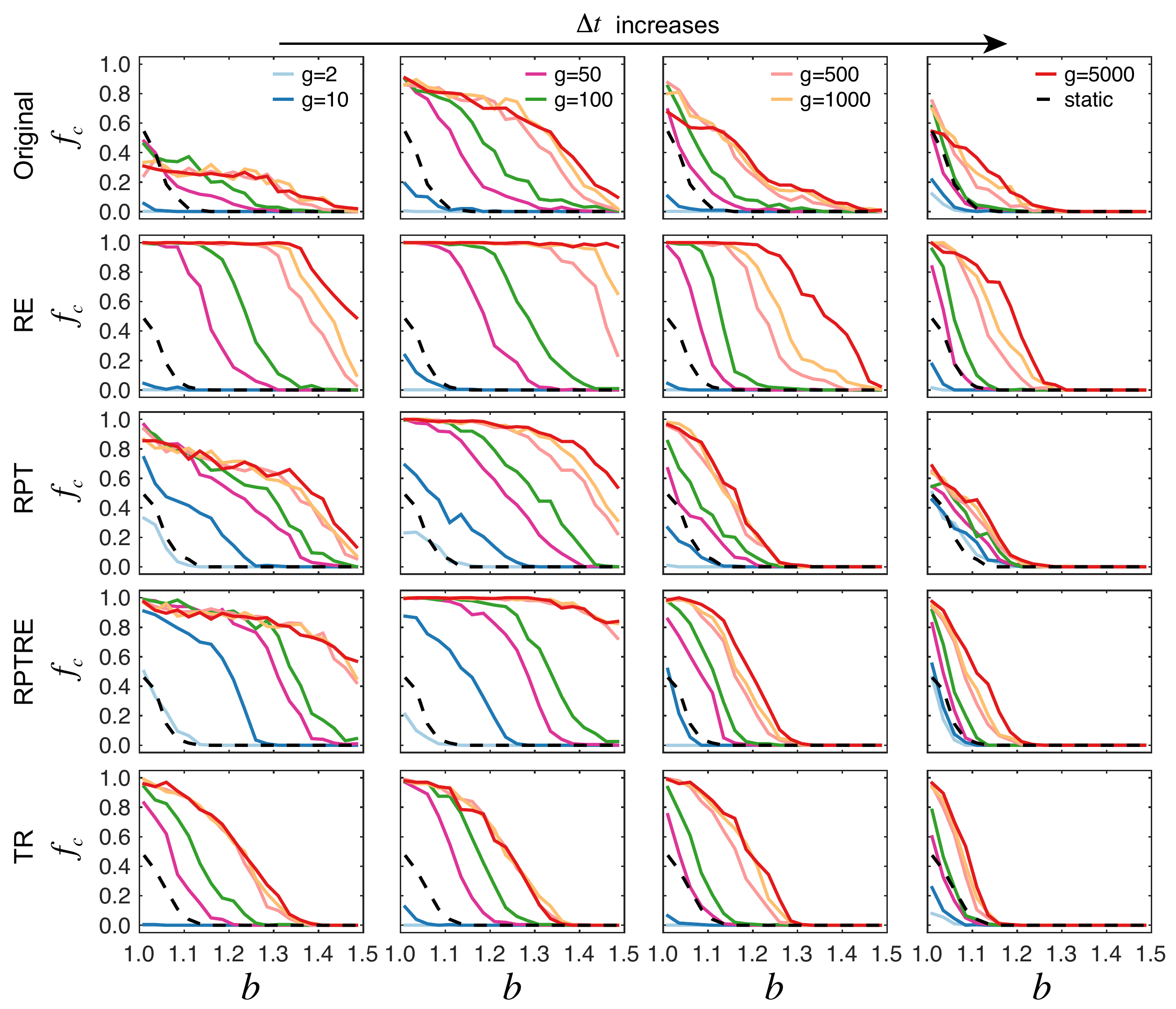}%
\caption{
\textbf{Evolution of cooperation on temporal networks generated from the original and randomized ACM conference dataset.}
For different null models, we 
show the stable fraction of cooperators $f_c$ as a function of the dilemma parameter $b$ for different $g$.
RE and TR have no effect on the correlations in temporal activity by construction, and hence have no effects on network temporality apparently.
RPT and RPTRE, on the other hand, destroy the temporal correlations between edges, thereby lowering the (too high) temporality of the system.
Actually for the temporal network where we run $g$ generation on each snapshot, the temporality of the underlying population structure is about $\mathcal{T}/g$.
Thus for small $g$ under RPT and RPTRE, $f_c$ is increased markedly relative to the original dataset, while 
for large $g$ the gains are more modest.
The above findings are also true for other datasets (see Figs.~\ref{fig_S_Stu2012} to~\ref{fig_S_Office2013}).
Overall, our results showing that temporal networks could facilitate the evolution of cooperation are robust even after the data is randomized.
Other parameters are the same as those in Fig.~\ref{fig_2}.
}
\label{fig_S_ht09}
\end{figure}

\begin{figure}[H]
\centering
\includegraphics[width=\textwidth]{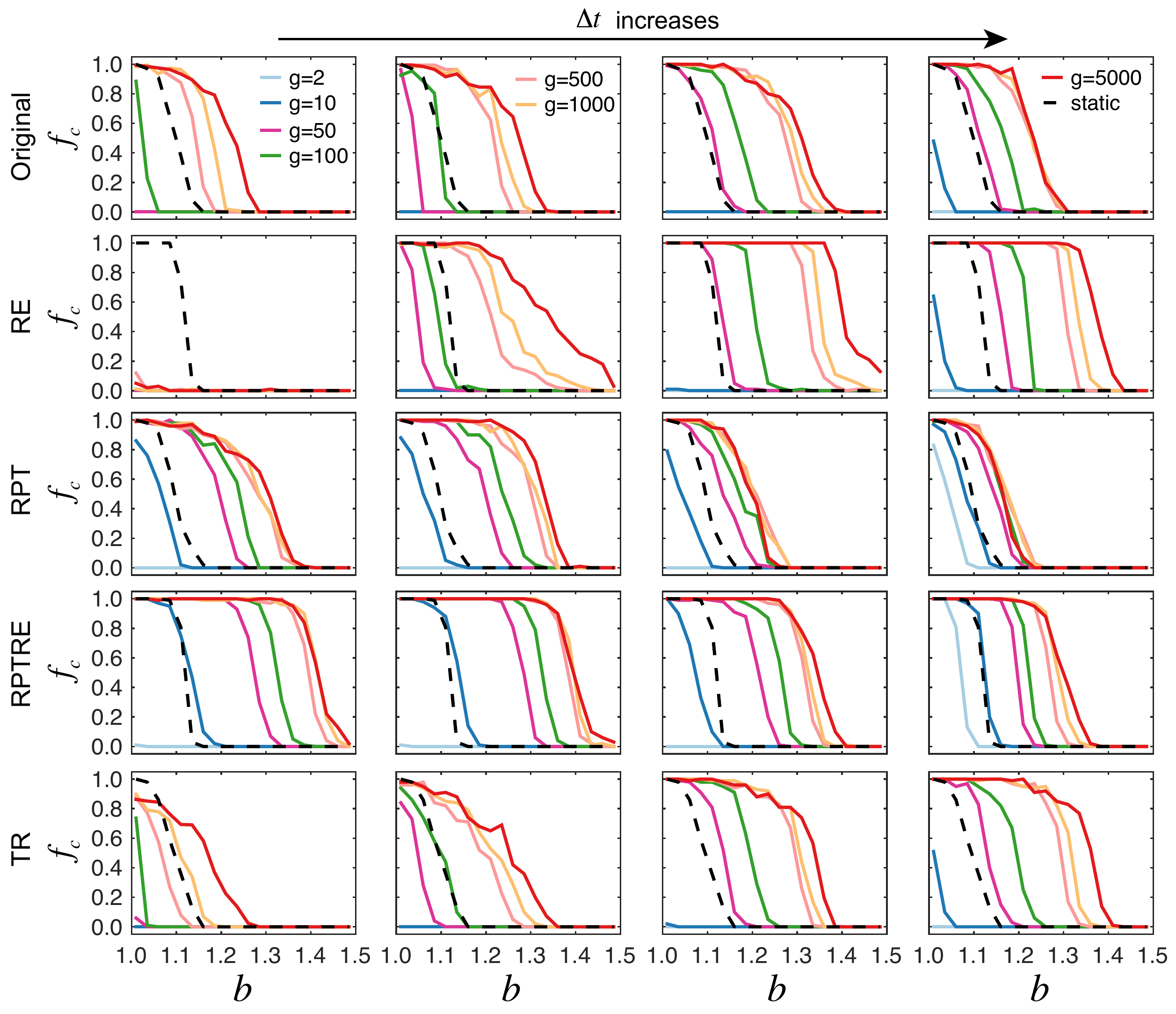}%
\caption{
\textbf{Evolution of cooperation on temporal networks generated from the original and randomized Student 2012 dataset.}
Note that when $\Delta t$ is small, 
Randomized Edges (RE) has the effect of breaking up the network structure within the (already sparse) snapshots, inhibiting cooperation.
Likewise, when $g$ is big, RPT fails to improve $f_c$ either owing to the small resulting temporality.
All parameters are the same as those in Fig.~\ref{fig_2}.
}
\label{fig_S_Stu2012}
\end{figure}

\begin{figure}[H]
\centering
\includegraphics[width=\textwidth]{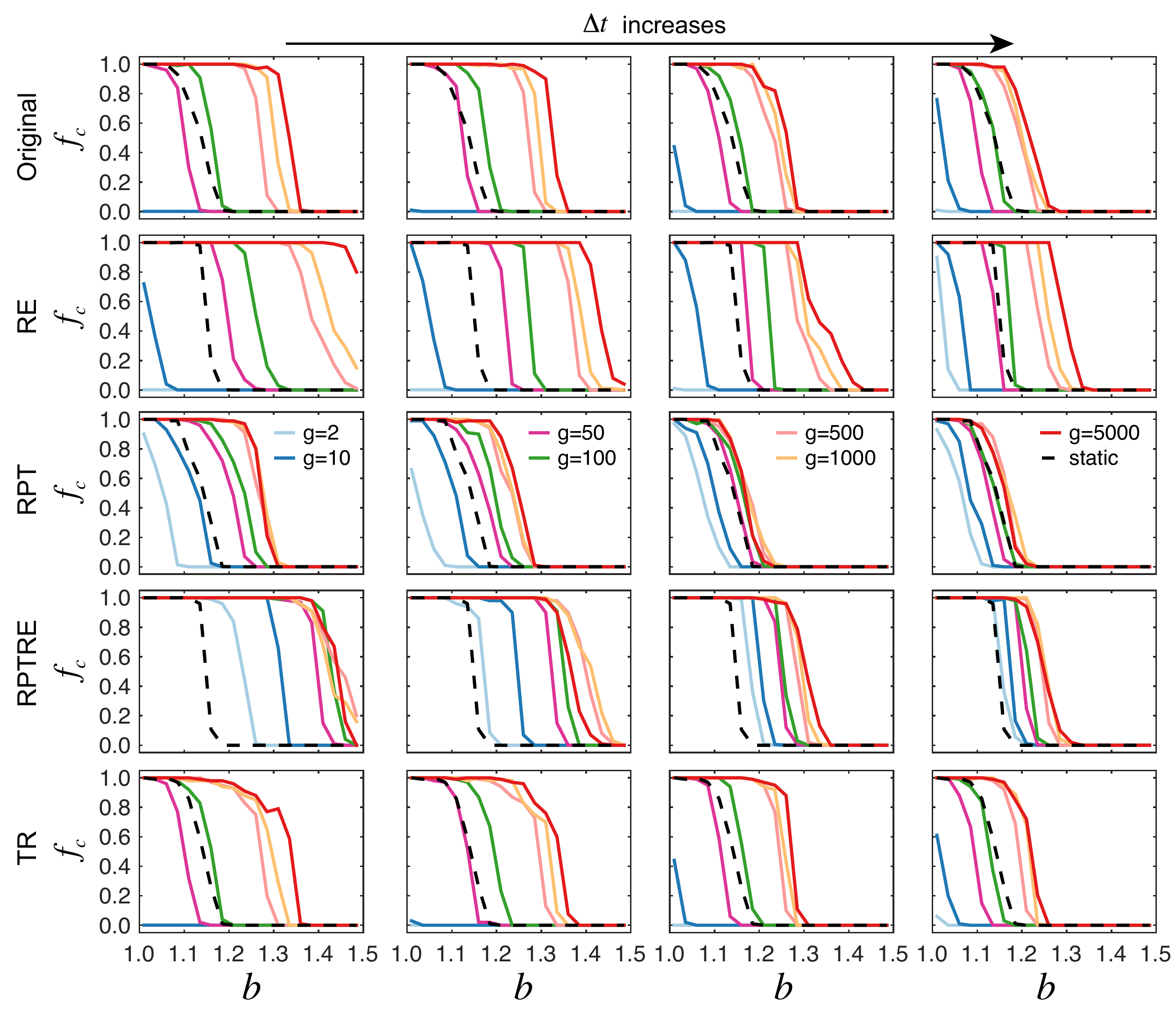}%
\caption{
\textbf{Evolution of cooperation on temporal networks generated from the original and randomized Student 2013 dataset.}
All parameters are the same as those in Fig.~\ref{fig_2}.
}
\label{fig_S_Stu2013}
\end{figure}

\begin{figure}[H]
\centering
\includegraphics[width=\textwidth]{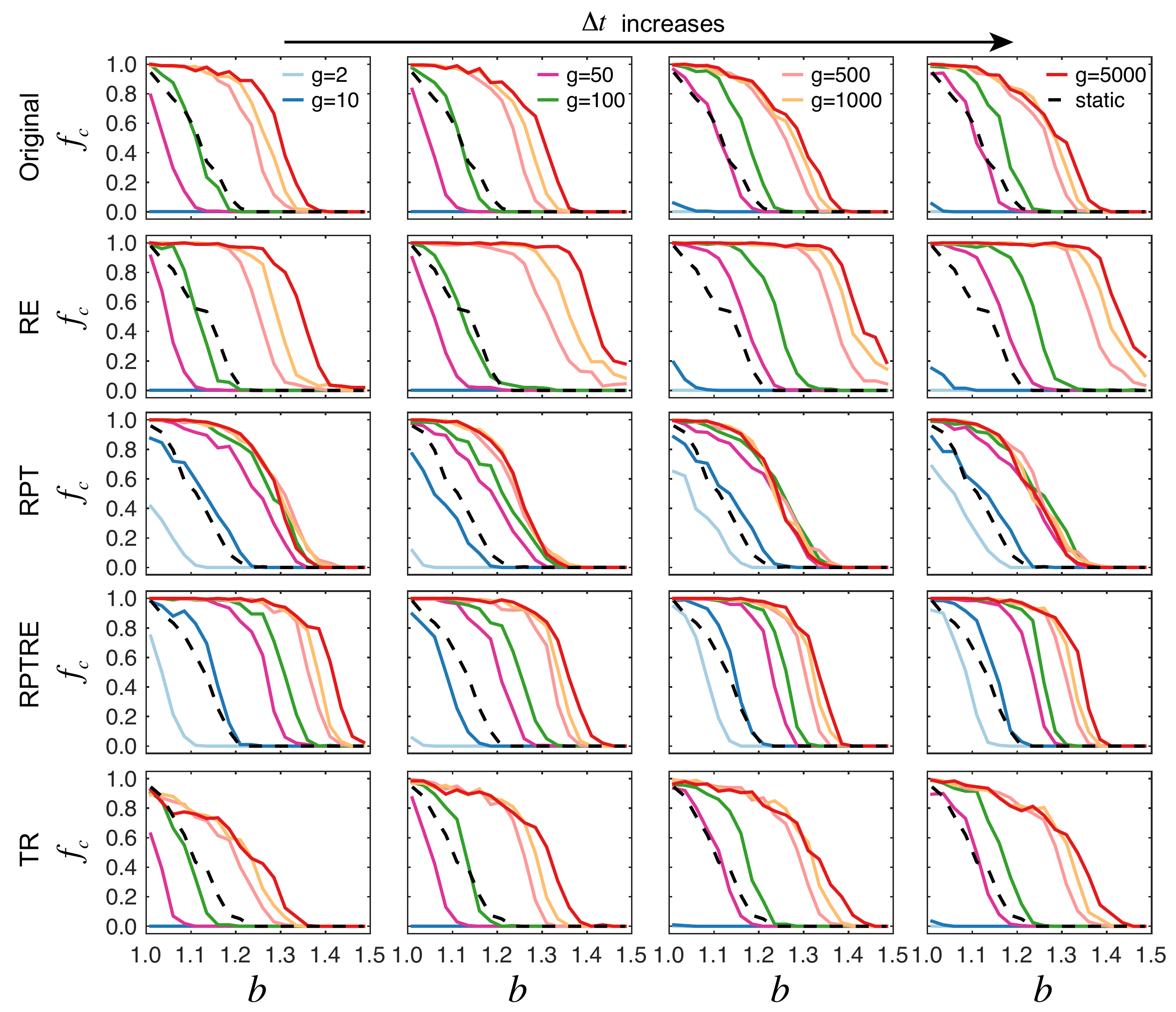}%
\caption{
\textbf{Evolution of cooperation on temporal networks generated from the original and randomized Office 2013 dataset.}
All parameters are the same as those in Fig.~\ref{fig_2}.
}
\label{fig_S_Office2013}
\end{figure}

\begin{figure}[H]
\centering
\includegraphics[width=\textwidth]{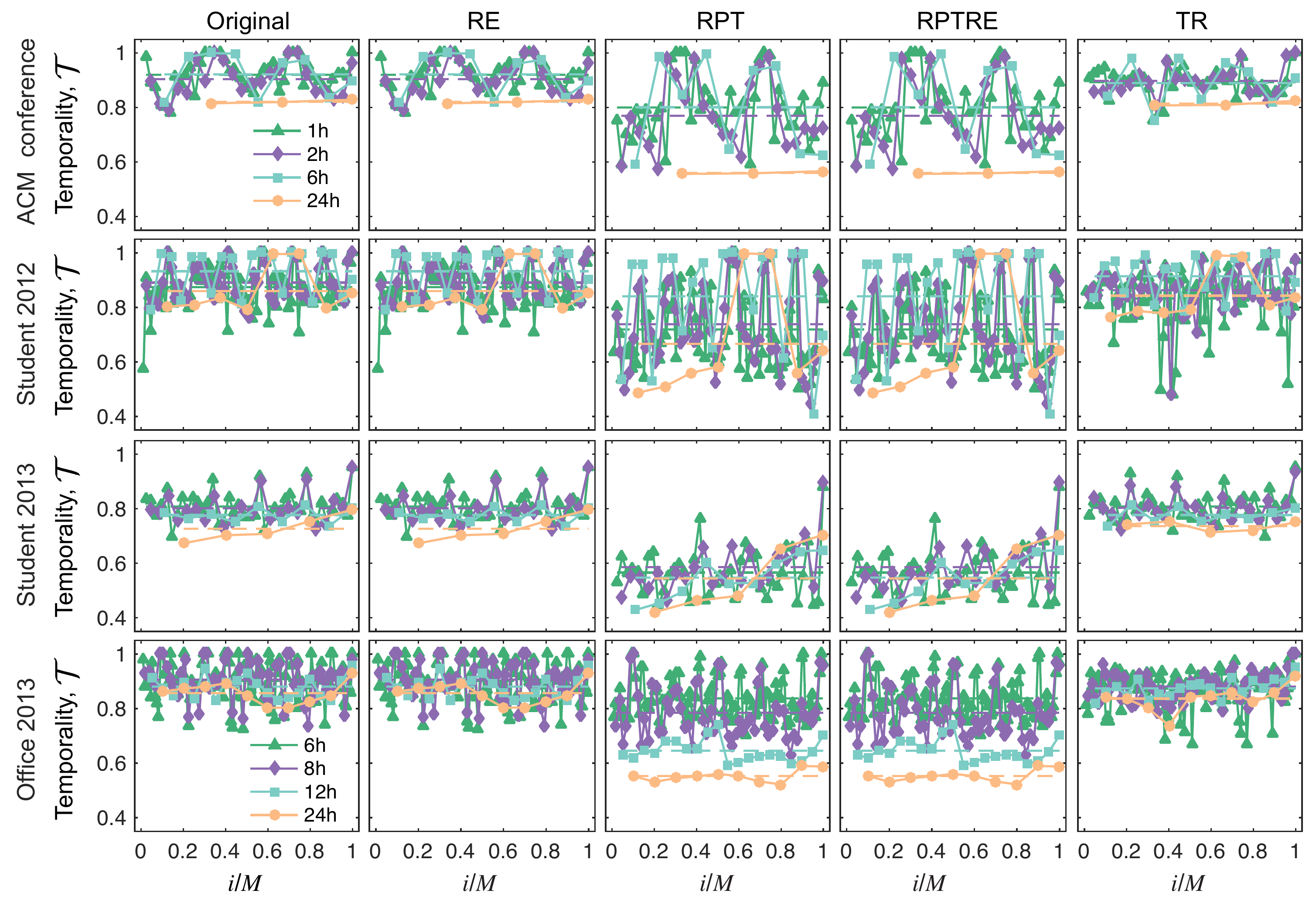}%
\caption{
\textbf{How temporality arises from differences between successive snapshots.}
Each time point shows the contribution to the temporality as defined in the main text made by each pair of snapshots $m$ and $m +1$. 
The total temporality $\mathcal{T}$ is the average of these contributions. 
Randomization methods that destroy temporal correlations in nodal activity (RPT, RPTRE) have the effect of lowering this average.
For every curve, we normalize the index of each snapshot under different $\Delta t$ by dividing the corresponding $\lceil T/ \Delta t \rceil$.
}
\label{fig_S_DetailT}
\end{figure}

\end{document}